\documentclass[twocolumn,superscriptaddress,prc,nofootinbib,floatfix]{revtex4-1}
\usepackage{graphicx}
\usepackage{color}
\usepackage{calc}
\usepackage{tikz}
\usepackage{bm}
\usepackage[version=3]{mhchem}
\usepackage{siunitx}
\usepackage{amsmath}
\usepackage[colorlinks = true,
            linkcolor = cyan,
            urlcolor  = blue,
            citecolor = red,
            anchorcolor = blue]{hyperref}
\usepackage{color}
\usepackage{amssymb}
\usepackage{amsthm}
\usepackage{textcomp}
\usepackage{float}
\usepackage{lineno}
\usetikzlibrary{matrix,chains,scopes,positioning,arrows,fit}
\newcommand{\poten}{$^{210}$Po}
\newcommand{\poeight}{$^{218}$Po}
\newcommand{\pofour}{$^{214}$Po}
\newcommand{\bifour}{$^{214}$Bi}
\newcommand{\rntwo}{$^{222}$Rn}
\newcommand{\pbten}{$^{210}$Pb}
\newcommand{\arnine}{$^{39}$Ar}
\newcommand{\LY}{light yield}
\newcommand{\fninety}{$f_{90}$}
\newcommand{\lyfactor}{$1.020 \pm 0.002$}
\begin{document}

%\preprint{APS/123-QED}

\title{Effect of Low Electric Fields on Alpha Scintillation Light Yield in Liquid Argon}

%\author{Chris Stanford}
%\affiliation{Department of Physics, Princeton University, Princeton, New Jersey 08544, USA}
\newcommand{\APC}{APC, Universit\'e Paris Diderot, CNRS/IN2P3, CEA/Irfu, Obs de Paris, USPC, Paris 75205, France}
\newcommand{\AQLNGS}{INFN Laboratori Nazionali del Gran Sasso, Assergi (AQ) 67100, Italy}
\newcommand{\AQGSSI}{Gran Sasso Science Institute, L'Aquila 67100, Italy}
\newcommand{\Augustana}{Physics Department, Augustana University, Sioux Falls, SD 57197, USA}
\newcommand{\Belgorod}{Radiation Physics Laboratory, Belgorod National Research University, Belgorod 308007, Russia}
\newcommand{\BHSU}{School of Natural Sciences, Black Hills State University, Spearfish, SD 57799, USA}
\newcommand{\BINP}{Budker Institute of Nuclear Physics, Novosibirsk 630090, Russia}
\newcommand{\BOINFN}{INFN Bologna, Bologna 40126, Italy}
\newcommand{\BOUniPHY}{Physics Department, Universit\`a degli Studi di Bologna, Bologna 40126, Italy}
\newcommand{\CAUniCHE}{Department of Mechanical, Chemical, and Materials Engineering, Universit\`a degli Studi, Cagliari 09042, Italy}
\newcommand{\CAINFN}{INFN Cagliari, Cagliari 09042, Italy}
\newcommand{\CAUniPHY}{Physics Department, Universit\`a degli Studi di Cagliari, Cagliari 09042, Italy}
\newcommand{\Campinas}{Physics Institute, Universidade Estadual de Campinas, Campinas 13083, Brazil}
\newcommand{\CentroFermi}{Museo della fisica e Centro studi e Ricerche Enrico Fermi, Roma 00184, Italy}
\newcommand{\CIEMAT}{CIEMAT, Centro de Investigaciones Energ\'eticas, Medioambientales y Tecnol\'ogicas, Madrid 28040, Spain}
\newcommand{\Cluj}{National Institute for R\&D of Isotopic and Molecular Technologies, Cluj-Napoca, 400293, Romania}
\newcommand{\CTLNS}{INFN Laboratori Nazionali del Sud, Catania 95123, Italy}
\newcommand{\ENUniCEE}{Civil and Environmental Engineering Department, Universit\`a degli Studi di Enna ``Kore'', Enna 94100, Italy}
\newcommand{\ETHZ}{Institute for Particle Physics, ETH Z\"urich, Z\"urich 8093, Switzerland}
\newcommand{\FNAL}{Fermi National Accelerator Laboratory, Batavia, IL 60510, USA}
\newcommand{\FortLewis}{Department of Physics and Engineering, Fort Lewis College, Durango, CO 81301, USA}
\newcommand{\GEUni}{Physics Department, Universit\`a degli Studi di Genova, Genova 16146, Italy}
\newcommand{\GEINFN}{INFN Genova, Genova 16146, Italy}
\newcommand{\GlenEllyn}{Glen Ellyn, Illinois 60137, USA}
\newcommand{\Hawaii}{Department of Physics and Astronomy, University of Hawai'i, Honolulu, HI 96822, USA}
\newcommand{\Houston}{Department of Physics, University of Houston, Houston, TX 77204, USA}
\newcommand{\IHEP}{Institute of High Energy Physics, Beijing 100049, China}
\newcommand{\INSTM}{Interuniversity Consortium for Science and Technology of Materials, Firenze 50121, Italy}
\newcommand{\IPHC}{IPHC, Universit\'e de Strasbourg, CNRS/IN2P3, Strasbourg 67037, France}
\newcommand{\JINR}{Joint Institute for Nuclear Research, Dubna 141980, Russia}
\newcommand{\Krakow}{M. Smoluchowski Institute of Physics, Jagiellonian University, 30-348 Krakow , Poland}
\newcommand{\Kurchatov}{National Research Centre Kurchatov Institute, Moscow 123182, Russia}
\newcommand{\LPNHE}{LPNHE, Universit\'e Pierre et Marie Curie, CNRS/IN2P3, Sorbonne Universit\'es, Paris 75252, France}
\newcommand{\MEPhI}{National Research Nuclear University MEPhI, Moscow 115409, Russia}
\newcommand{\MIBIINFN}{INFN Milano Bicocca, Milano 20126, Italy}
\newcommand{\MIINFN}{INFN Milano, Milano 20133, Italy}
\newcommand{\MIPoliICA}{Civil and Environmental Engineering Department, Politecnico di Milano, Milano 20133, Italy}
\newcommand{\MIPoliCHE}{Chemistry, Materials and Chemical Engineering Department ``G.~Natta", Politecnico di Milano, Milano 20133, Italy}
\newcommand{\MIPoliEIB}{Electronics, Information, and Bioengineering Department, Politecnico di Milano, Milano 20133, Italy}
\newcommand{\MIPoliENE}{Energy Department, Politecnico di Milano, Milano 20133, Italy}
\newcommand{\MIUni}{Physics Department, Universit\`a degli Studi di Milano, Milano 20133, Italy}
\newcommand{\MSU}{Skobeltsyn Institute of Nuclear Physics, Lomonosov Moscow State University, Moscow 119991, Russia}
\newcommand{\NAINFN}{INFN Napoli, Napoli 80126, Italy}
\newcommand{\NAUniPHY}{Physics Department, Universit\`a degli Studi ``Federico II'' di Napoli, Napoli 80126, Italy}
\newcommand{\NAUniCHE}{Chemical, Materials, and Industrial Production Engineering Department, Universit\`a degli Studi ``Federico II'' di Napoli, Napoli 80126, Italy}
\newcommand{\NSU}{Novosibirsk State University, Novosibirsk 630090, Russia}
\newcommand{\Petersburg}{Saint Petersburg Nuclear Physics Institute, Gatchina 188350, Russia}
\newcommand{\PGUniCBB}{Chemistry, Biology and Biotechnology Department, Universit\`a degli Studi di Perugia, Perugia 06123, Italy}
\newcommand{\PGINFN}{INFN Perugia, Perugia 06123, Italy}
\newcommand{\PIINFN}{INFN Pisa, Pisa 56127, Italy}
\newcommand{\PIUniPHY}{Physics Department, Universit\`a degli Studi di Pisa, Pisa 56127, Italy}
\newcommand{\PNNL}{Pacific Northwest National Laboratory, Richland, WA 99352, USA}
\newcommand{\Princeton}{Physics Department, Princeton University, Princeton, NJ 08544, USA}
\newcommand{\RMTreINFN}{INFN Roma Tre, Roma 00146, Italy}
\newcommand{\RMTreUni}{Mathematics and Physics Department, Universit\`a degli Studi Roma Tre, Roma 00146, Italy}
\newcommand{\RMUnoINFN}{INFN Sezione di Roma, Roma 00185, Italy}
\newcommand{\RMUnoUni}{Physics Department, Sapienza Universit\`a di Roma, Roma 00185, Italy}
\newcommand{\SSUniCHP}{Chemistry and Pharmacy Department, Universit\`a degli Studi di Sassari, Sassari 07100, Italy}
\newcommand{\Temple}{Physics Department, Temple University, Philadelphia, PA 19122, USA}
\newcommand{\TNFBK}{Fondazione Bruno Kessler, Povo 38123, Italy}
\newcommand{\TNTIFPA}{Trento Institute for Fundamental Physics and Applications, Povo 38123, Italy}
\newcommand{\TNUni}{Physics Department, Universit\`a degli Studi di Trento, Povo 38123, Italy}
\newcommand{\TOINFN}{INFN Torino, Torino 10125, Italy}
\newcommand{\TOUni}{Physics Department, Universit\`a degli Studi di Torino, Torino 10125, Italy}
\newcommand{\UCDavis}{Department of Physics, University of California, Davis, CA 95616, USA}
\newcommand{\UCLA}{Physics and Astronomy Department, University of California, Los Angeles, CA 90095, USA}
\newcommand{\UMass}{Amherst Center for Fundamental Interactions and Physics Department, University of Massachusetts, Amherst, MA 01003, USA}
\newcommand{\UOC}{Department of Chemistry, University of Crete, P.O. Box 2208, 71003 Heraklion, Crete, Greece}
\newcommand{\USP}{Instituto de F\'isica, Universidade de S\~ao Paulo, S\~ao Paulo 05508-090, Brazil}
\newcommand{\VTech}{Virginia Tech, Blacksburg, VA 24061, USA}

\renewcommand{\thefootnote}{\fnsymbol{footnote}}
\newcommand\blfootnote[1]{%
  \begingroup
  \renewcommand\thefootnote{}\footnote{#1}%
  \addtocounter{footnote}{-1}%
  \endgroup
}
\blfootnote{* Corresponding author: jcjs@princeton.edu (C.~Stanford)}

\author{P.~Agnes}\affiliation{\APC}
\author{I.~F.~M.~Albuquerque}\affiliation{\USP}
\author{T.~Alexander}\affiliation{\PNNL}
\author{A.~K.~Alton}\affiliation{\Augustana}
\author{D.~M.~Asner}\affiliation{\PNNL}
\author{H.~O.~Back}\affiliation{\PNNL}
\author{B.~Baldin}\affiliation{\FNAL}
\author{K.~Biery}\affiliation{\FNAL}
\author{V.~Bocci}\affiliation{\RMUnoINFN}
\author{G.~Bonfini}\affiliation{\AQLNGS}
% ...
\collaboration{The DarkSide Collaboration}

\date{\today}

\begin{abstract}

Measurements were made of scintillation light yield of alpha particles from the \rntwo\ decay chain within the DarkSide-50 liquid argon time projection chamber. The light yield was found to increase as the applied electric field increased, with alphas in a 200\,V/cm electric field exhibiting a $\sim$2\% increase in light yield compared to alphas in no field. 

\end{abstract}

\maketitle

%\tableofcontents

%\linenumbers


DarkSide-50 is a dark matter experiment searching for Weakly Interacting Massive Particles (WIMPs) using a two-phase Time Projection Chamber (TPC) containing a liquid argon (LAr) target mass of roughly 50\,kg. When ionizing radiation interacts with the LAr, two signals are produced in the TPC. First, the argon undergoes deexcitation and recombination, and produces primary scintillation (S1). Second, ionization electrons are drifted to the surface of the liquid by an electric field (drift field) and extracted by a second electric field (extraction field) to produce secondary scintillation in a layer of argon gas above the liquid (S2). Both the S1 and S2 signals are detected by arrays of 19 photomultiplier tubes (PMTs) at the top and bottom of the TPC. The signals are measured in terms of photoelectrons (PE), calibrated with a low-intensity laser. A more detailed description of the detector can be found in \cite{ref:ds50day,ref:ds70day}.

The S1 light yield (hereafter referred to as simply \LY) for an interaction is defined as $LY=S1/Q$, where Q is the energy deposited in the argon during the interaction.

For beta particles, the presence of an electric field reduces the \LY~\cite{ref:PhysRevD.91.092007}. This is typically explained as a result of the field pulling away electrons from the positive argon ions, resulting in less recombination and thus less primary scintillation. DarkSide-50 reports a beta light yield of 7.9\,PE/keV for 0\,V/cm drift field and 7.0\,PE/keV for 200\,V/cm drift field~\cite{ref:ds50day}.

However, this relationship does not hold true for all types of ionizing particles. The \LY\ of liquid argon as a function of drift field has been measured for many different forms of radiation, including neutrons~\cite{ref:PhysRevD.91.092007}, fission fragments~\cite{ref:PhysRevA.35.3956}, alpha particles~\cite{ref:PhysRevA.35.3956,ref:PhysRevB.46.540}, and helium ions~\cite{ref:PhysRevB.46.540}. Alphas and helium ions were found to behave differently from the rest; the \LY\ increased as the drift field increased, until it reached a peak when the drift field was on the order of 1\,kV/cm, after which the \LY\ began to decrease.

\begin{figure}[tp]
%\makebox[0.3\textwidth][c]{\input{rnchain.tex}}
\centering
\includegraphics[width=0.48\textwidth]{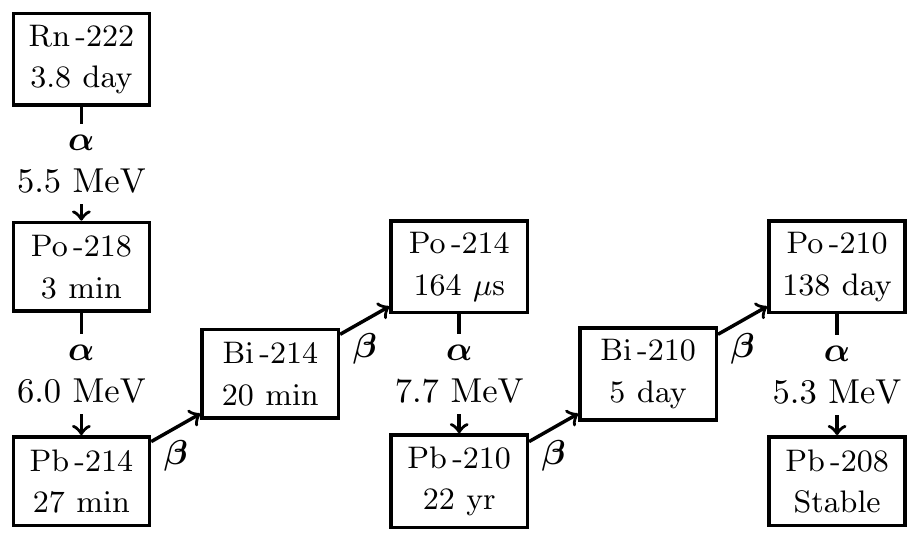}
 \caption{The Rn-222 decay chain with the energy of the decay product and the half-life of each state. The energies listed are for the primary (\textgreater 99.9\%) alpha decay modes. }
 \label{fig:rnchain}
\end{figure}

The purpose of this paper is to report similar findings of the \LY\ behavior for alpha induced scintillation in DarkSide-50. A sample of alpha decays is obtained through the detection of naturally occurring radon daughters present in the liquid (Figure  \ref{fig:rnchain}). Despite its low level of radon contamination ($\mathrm{\mu Bq/kg}$), DarkSide-50 has collected enough statistics to resolve the peaks in the \rntwo\ decay chain spectrum.

For the dark matter search, DarkSide-50 operates with a drift field of 200\,V/cm. However, a substantial amount of data has been collected with drift fields in a range from 0 to 200\,V/cm for calibration purposes, and these data sets provide us with a means of comparing the alpha \LY\ at different field strengths.

Alphas produce S1 signals (``pulses") that are very large ($>$10000 PE), and they saturate the CAEN V1720 digitizers used for DarkSide-50's dark matter search. To recover these interactions, CAEN V1724 digitizers receiving a lower-gain signal from the PMTs were used in parallel with the V1720s~\cite{ref:v1720,ref:v1724}. The V1724 pulse integrals were converted to units of PE using a conversion factor derived from a linear mapping of V1724 to V1720 pulse integrals at lower, unsaturated levels.

A correction was applied to S1 that depends on the vertical ($z$) position of each alpha event within the TPC to account for a $z$-dependence of the light collection efficiency within the detector. The $z$-position of an event is typically measured using drift time (the time between S1 and S2), but in this case, where a comparison is being made to data taken with no field, drift time could not be used. A parameter called top-bottom asymmetry (TBA) was used instead. It is defined as:

$$\text{TBA} = \frac{\text{S1 from top PMTs - S1 from bottom PMTs}}{\text{S1 from all PMTs}}$$

\begin{figure}[tp]
\centering
\includegraphics[width=0.48\textwidth]{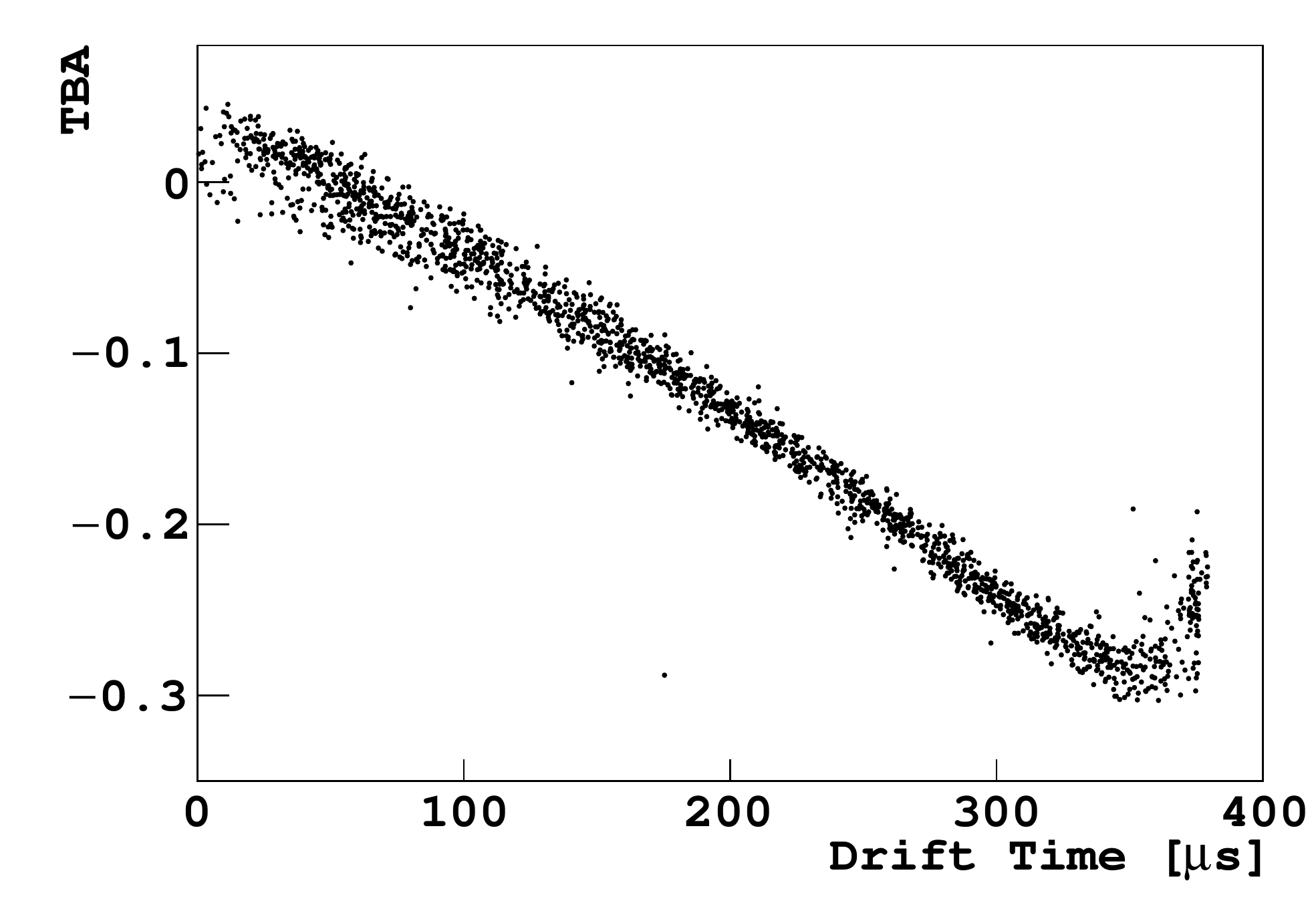}
\caption{Top-bottom asymmetry (TBA) vs. drift time for Po-218 events in a drift field of 200\,V/cm. TBA is used as a measure of an event's z-position to make a correction to the energy. See the text for an explanation of the feature at high drift time.}
\label{fig:tbavstdrift}
\end{figure}

A scatter plot of TBA and drift time in a 200\,V/cm drift field can be seen in Figure \ref{fig:tbavstdrift}. The feature at high drift time is a result of the geometry of the detector; events that are at the maximum drift time (the bottom surface of the TPC) have a TBA that is heavily biased depending on whether the event occurs above a PMT or above the PTFE reflector separating the PMTs. Events occurring directly above the reflector have more light reflecting into the top PMT array, biasing the TBA upward. There is no equivalent feature at low drift time because the liquid is separated from the upper surface of the TPC by a layer of argon gas. This effect introduces a small systematic error into the TBA correction. Since systematic errors from the TBA correction do not depend on the drift field, we eliminate these errors in the final result by presenting the \LY\ at each drift field relative to the 0\,V/cm \LY.

The TBA distribution is not symmetric about 0 due to partial internal reflection from the liquid-gas interface at the top of the TPC.

\begin{figure}[tp]
\centering
\includegraphics[width=0.48\textwidth]{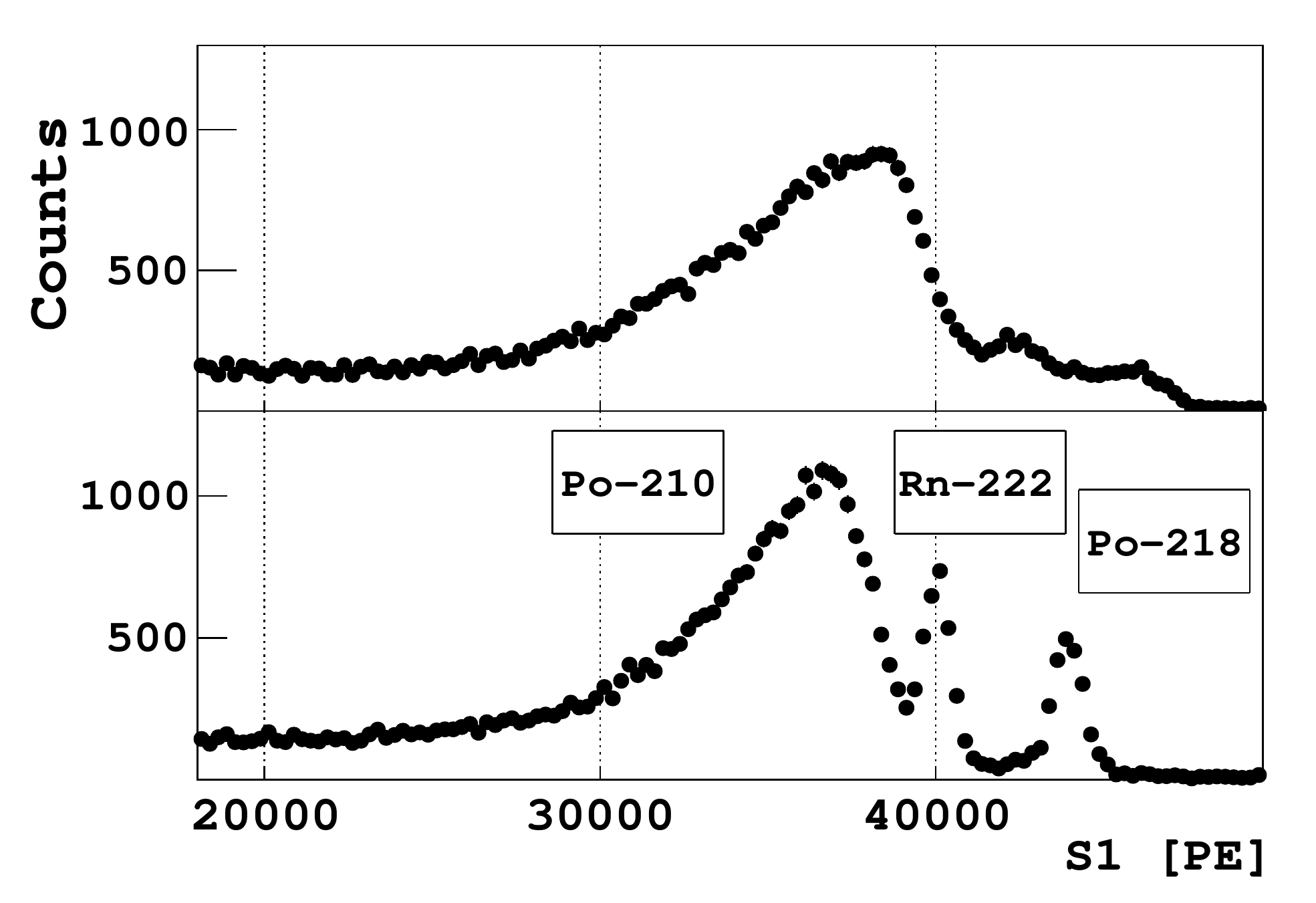}
\caption{The alpha energy spectrum taken with a 200\,V/cm drift field. \textbf{Top:} The spectrum after incorporating low-gain digitizer data to reconstruct the saturated pulse. \textbf{Bottom:} The top spectrum after correcting for the event's $z$-position within the detector, as determined by the top-bottom asymmetry.}
\label{fig:corrections}
\end{figure}

The S1 correction factor for a given TBA is derived from a calibration using the endpoint energy of the \arnine\ spectrum. The magnitude of the correction ranges from roughly $-10$\% to $+10$\% of S1.

The alpha spectrum after the TBA correction is shown in Figure \ref{fig:corrections}. The three peaks identifiable in the final spectrum are labelled based on their relative energies, from left to right, as \poten, \rntwo, and \poeight. The long tail toward lower energies in the \poten\ peak is attributed to the origin of the \poten\ being beneath the thin layer of wavelength shifter on the surface of the detector. This causes the alpha to lose a fraction of its energy depending on its angle of ejection from the wall. \poten\ is not expected to be in secular equilibrium with \rntwo\ and \poeight\ because the chain is separated by \pbten. \pbten\ has a 22yr half-life, and can be directly introduced to the detector through various means, such as the purified water used to clean the detector surfaces~\cite{ref:water}. The alphas from \pofour\ normally fall within the data acquisition window or dead time from the immediately preceding ($\tau=237$\,$\mu$s) beta from \bifour\, so these alphas are not tagged as S1 and do not appear in the spectrum.

\begin{figure}[tp]
\includegraphics[width=0.48\textwidth]{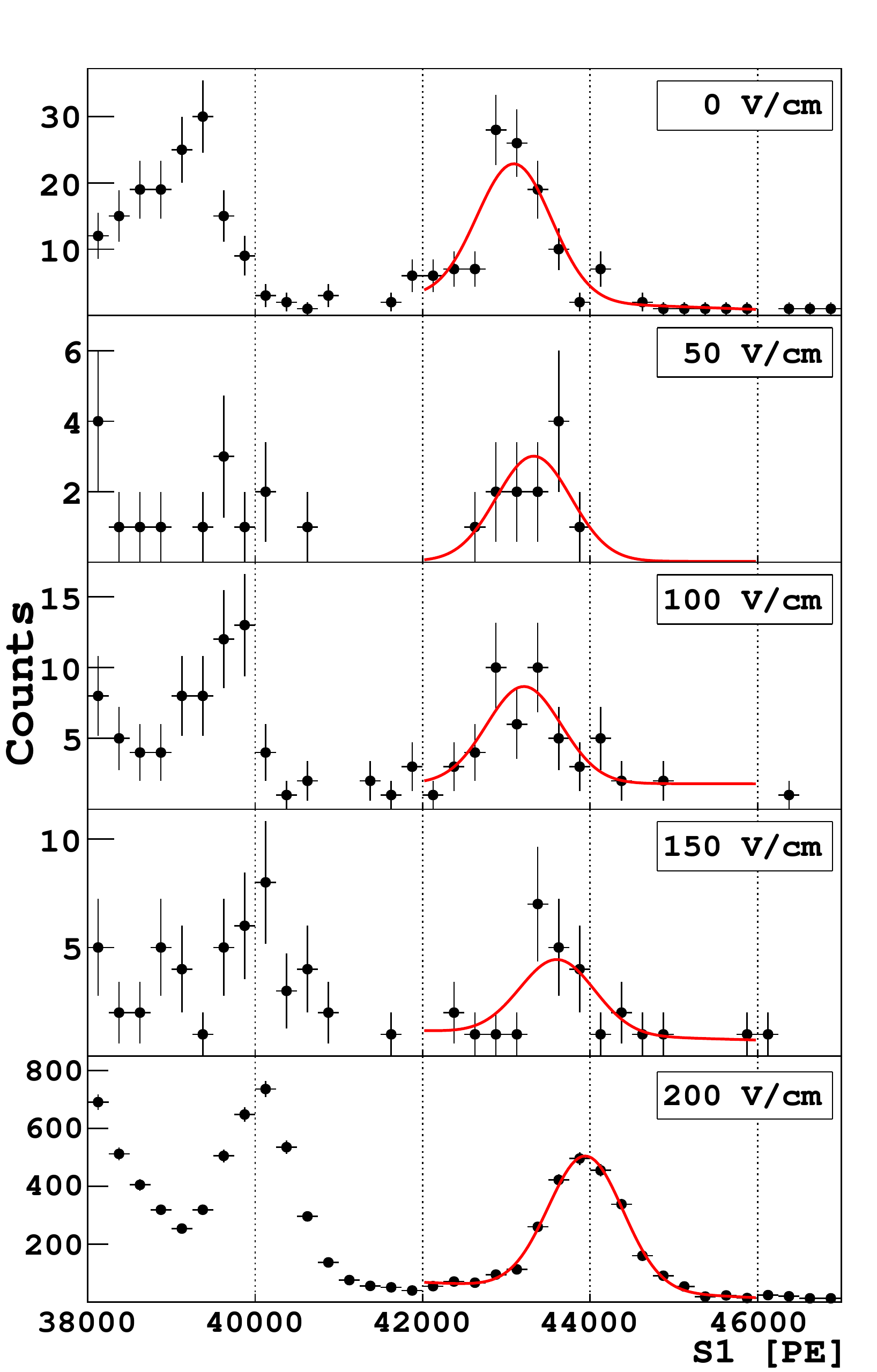}
\caption{The alpha energy spectrum at different drift fields. The \poeight\ peak, fit here with a Gaussian+linear function, noticeably increases in energy as the strength of the drift field increases. A coherent shift in the Rn-222 peak is visible on the left, although no attempt was made to fit this region.}
\label{fig:comparison_main}
\end{figure}

Cosmogenic backgrounds, such as muons, were tagged using a liquid scintillator veto and water Cherenkov detector surrounding the TPC. Events occurring in the 2\,s following a muon detection were cut. Further details about the DarkSide-50 veto detectors can be found in~\cite{ref:veto}. 

The only other background in the alpha energy range are events where the S1 pulse was missed due to dead time in the data acquisition and the data acquisition triggered on the subsequent S2, which commonly measure in the tens of thousands of PE. These events can be easily removed with a pulse-shape cut because S2 pulses have a longer rise-time than S1 pulses. The pulse-shape parameter used for the cut is called \fninety\ and is defined as the integral of the first 90\,ns of the pulse divided by the integral of the entire pulse. The \fninety\ is 0.6-0.8 for alpha S1s\footnote{The \fninety\ for alphas is determined using a tagging method that identifies alphas from \pofour\ based on its coincidence with \bifour.} and 0-0.1 for S2s. Events with \fninety$<0.5$ were cut, providing a clean sample of alpha S1s.

The light yield at each drift field is calculated with a fit to the \poeight\ peak. No conclusions were drawn from the larger \poten\ peak because the \poten\ alphas are coming from decays on the surfaces, and the electric field along the surfaces is not as well understood as in the bulk. The \rntwo\ peak was not used due to leakage from the broad \poten\ peak.

The alpha spectra for drift fields of 0, 50, 100, 150, and 200\,V/cm can be found in Figure \ref{fig:comparison_main}. The histograms are defined with a bin width of 250\,PE. The \poeight\ peak is fit with a Gaussian+linear function over the domain (42000 PE, 46000 PE). To assist in the fit of the intermediate drift fields with lower statistics, an assumption was made that the peak width remains constant over the relatively small increase in mean. Then, a binned log likelihood fitting method was used across all the histograms simultaneously, with the peak width constrained to be the same while the mean was allowed to vary.

\begin{figure}[tp]
\includegraphics[width=0.48\textwidth]{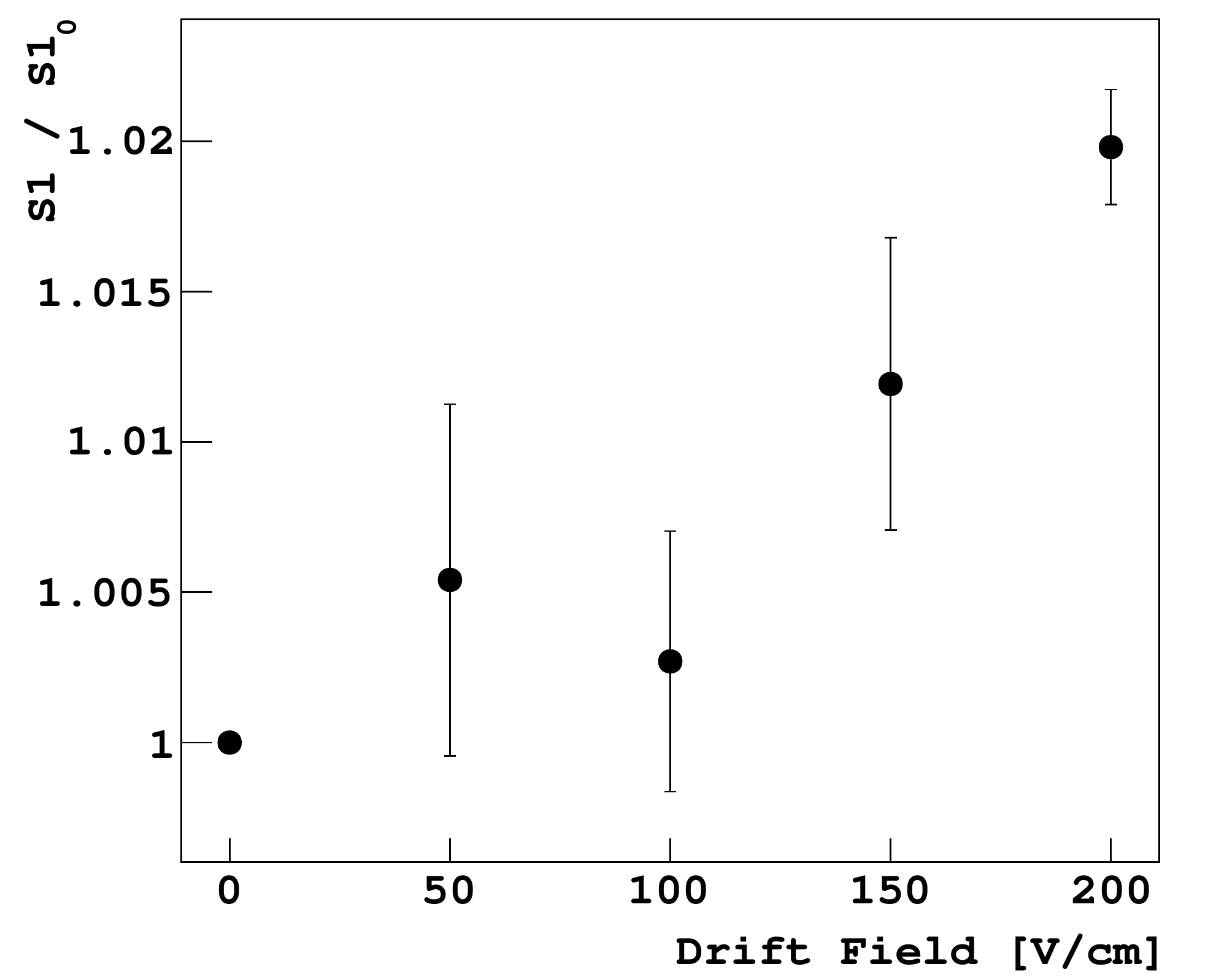}
\caption{\poeight\ alpha (6.0\,MeV) S1 light yield at various drift field strengths relative to 0\,V/cm. Statistical errors only.}
\label{fig:linear_fit}
\end{figure}

The results show that alpha \LY\ in DarkSide-50 increases as the drift field increases from 0\,V/cm to 200\,V/cm. The \LY\ of \poeight\ alphas at 200\,V/cm is \lyfactor\ times the \LY\ at 0\,V/cm. This ratio for each drift field is plotted in Figure \ref{fig:linear_fit}. The errors are statistical only, with the 0 V/cm point equal to 1 by definition.

The systematic errors from the choice of binning and fitting region were studied and found to be much smaller than the statistical errors.

In the previously mentioned study by Hitachi et al.~\cite{ref:PhysRevB.46.540}, a similar effect was observed with \poten\ alphas (5.3\,MeV). Although their study measures the dependence over a larger range (up to 6\,kV/cm), their first data point is not until 540\,V/cm (with S1/S1$_0=1.03$). Their following points at higher electric field strengths show a decreasing \LY. Our study, in addition to confirming the phenomenon for \poeight\ alphas, fills in the valuable region at low field strengths where the light yield is still increasing.

The cause of the increasing alpha \LY\ is not fully understood. As a possible explanation, Hitachi et al. suggest that the electric field may help drift the electrons out of the dense ionization tracks characteristic of alphas and helium ions, and allow some of the recombination to occur in regions of lower ionization density where non-radiative quenching processes have a reduced effect.

\section*{ACKNOWLEDGMENTS}
The DarkSide-50 Collaboration would like to thank LNGS laboratory and its staff for invaluable technical and logistical support. This report is based upon work supported by the US NSF (Grants PHY-0919363, PHY-1004072, PHY-1004054, PHY-1242585, PHY-1314483, PHY-1314507 and associated collaborative grants; Grants PHY-1211308, PHY-1606912, and PHY-1455351), the Italian Istituto Nazionale di Fisica Nucleare (INFN), the US DOE (Contract Nos. DE-FG02-91ER40671 and DE-AC02-07CH11359), the Polish NCN (Grant UMO-2014/15/B/ST2/02561), and the Russian Science Foundation Grant No 16-12-10369. We thank the staff of the Fermilab Particle Physics, Scientific and Core Computing Divisions for their support. We acknowledge the financial support from the UnivEarthS Labex program of Sorbonne Paris Cit\'e (ANR-10-LABX-0023 and ANR-11-IDEX-0005-02) and from the S\~ao Paulo Research Foundation
(FAPESP).

\bibliographystyle{apsrev}
\bibliography{bibliography}

\end{document}